\documentclass[12pt,letterpaper]{article}

\newcommand{\required}[1]{\section*{\centering{#1}}}                    %%
\usepackage{amsmath}
\usepackage{amssymb}
\usepackage{xcolor}
\usepackage[numbers,sort&compress]{natbib}
\usepackage{multirow}
\usepackage{multicol}
\usepackage{url}
\usepackage{graphicx}
\usepackage{subfig}
\usepackage{wrapfig}
\usepackage[justification=centering]{caption}
\usepackage{comment}

\begin{document}

\setcounter{page}{1}

\include{Summary}
 \setcounter{page}{1}
%%%%%%%%% PROPOSAL -- 15 pages (including Prior NSF Support)

\required{Understanding and Developing Equitable and Fair Transportation Systems}
\centerline{PI: Weizi Li} 
\centerline{University of Memphis} 
\centerline{E-mail: wli@memphis.edu}

\section{Introduction}
\label{sec:intro}
The transportation system is an interplay between infrastructure, vehicles, and policy. During the past century, the rapid expansion of the road network, blended with increasing vehicle production and mobility demands, has been stressing the system's capacity and resulting in annual expenditures of \$90 billion (traffic congestion)~\cite{scorecard2015texas}, \$900 billion (traffic accidents)~\cite{blincoe2015economic}, and \$100 billion (infrastructure maintenance). In order to alleviate these costs while providing passengers with safe and efficient travel experiences, we need to better design and plan our transportation system. 

To start with, the design of our road network isn't the most efficient. Topologically, roads are by large linearly coupled and lack fail-safe redundancy, which is commonly found in other safety-critical systems~\cite{perrow2011normal}. One incident involving two vehicles can completely disable streams of roads and cause traffic jams to spread for miles. Geographically, our road network not only embeds but likely facilitates inequality:  roads and bridges are found to better connect affluent sectors, while excluding the poor~\cite{bullard2004highway}. In addition, our road network is constantly under the stressors of extreme conditions such as wildfires, hurricanes, tornadoes, floods, which further challenge its resilience and threaten essential mobility activities. 

While exhausting transport policies and control methods~\cite{orosz2010traffic} to contain the increasing costs of the transportation system, technological advancements such as connected and autonomous vehicles (CAVs) and novel operation modes such as shared economy have offered new opportunities. However, many questions remain. First, what is the relationship between the road network, community development, demographics, and mobility behaviors? Second, by leveraging the insights from studying the first question, can we better plan, coordinate, and optimize vehicles in different modalities such as human-driven and autonomous to construct safe, efficient, and resilient traffic flows? Third, how can we build an intelligent transportation system to promote equity and fairness in our community development?     

This proposal is the first step towards answering these questions. Specifically, we plan to leverage the unprecedented traffic data as a result of various lockdown policies implemented in 2020--2021. By comparing pre-lockdown and post-lockdown traffic data, we aim to reveal the connections between road network features, demographic factors, and traffic dynamics including traffic accidents and congestion. Next, we will integrate the obtained insights with advanced simulation and machine learning techniques to explore the potential of using CAVs to build safer, more resilient, and more equitable intelligent transportation systems.   
 
\section{Background and Preliminary Study}
\label{sec:related}
As a preliminary study, we investigate the influence of lockdown, i.e., reduced traffic flows, on traffic accidents. In this section, we will briefly introduce our findings (unpublished), upon which the proposed research will be conducted.

\subsection*{Demographic Inequality}
\begin{figure*}[ht]
\centering
\includegraphics[width=\linewidth]{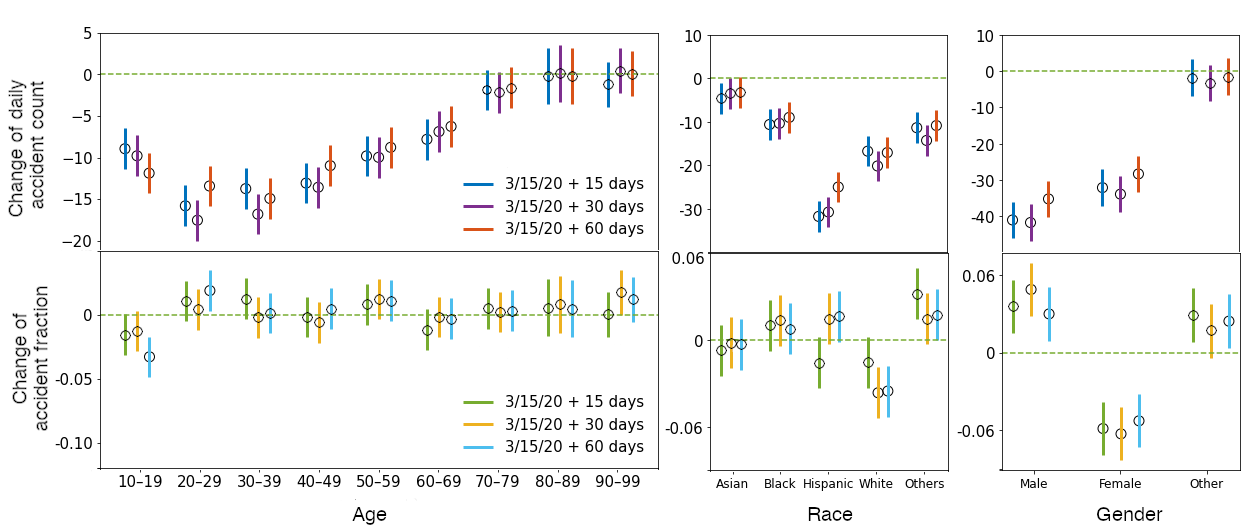}
\vspace*{-.5em}
\caption{\footnotesize{Change of daily accidents after lockdown across various age, race, and gender groups. Top: Change of daily accident count in each demographic group. Bottom: Change of daily accident fraction in each demographic group. 95\% confidence intervals are shown. Also shown are estimates from three time windows, namely 15 days , 30 days, and 60 days, before and after lockdown.}}
\label{fig:demographics}
\end{figure*}

In Fig.~\ref{fig:demographics}, we show the change of daily traffic accidents in different demographic groups before and after lockdown. Age, race, and gender are considered. The top row presents the changes in daily accident counts, which are mostly negative, meaning the number of accidents decreases across nearly all groups. However, the share of the accidents of each group (i.e., the fraction of accidents of one group divided by the total number of accidents) shows a different pattern at the bottom row. This observation demonstrates the disproportional impact of the pandemic on different demographic groups. To assess the robustness of the results, we consider three time windows: 15 days, 30 days, and 60 days, before and after lockdown. We have also removed seasonal shift in the changes and provided 95\% confidence interval for statistical significance. 

Regarding age, all groups except \emph{70--79}, \emph{80--89}, and \emph{90--99} have significant reductions in daily accident counts. For groups older than 70, the changes are insignificant, which may due to the fact that seniors in general travel less and hence are impacted less by the mobility change. The groups \emph{20--29} and \emph{30--39} have the largest decrease in accident counts, but their shares do not change significantly. Only the group \emph{10--19} has a significant decrease of 3.2\% in its share. Overall, the pandemic does not seem to have a largely biased impact across all age groups. Regarding race, all groups experience a significant reduction in accident counts, among which \emph{Hispanic} has the largest decrease followed by an increase after 15 days. In comparison, \emph{White} has a significant reduction in both the number of accidents and fraction of accidents. Lastly, for gender, both \emph{Male} and \emph{Female} have significant reductions in the number of accidents. However, in terms of the share of accidents, \emph{Male} has increased about 4\%, while \emph{Female} has decreased about 5\%. In sum, the distribution of accidents has shifted its mass towards \emph{Hispanic} and \emph{Male}.

\vspace*{-1em}

\subsection*{Spatial Irregularity}
% Fig.~\ref{fig:temporal_spatial}A shows the change of accidents at different hours of a day. The hours between 06:00 and 22:00 have seen significant decreases in accident counts. Especially during the morning (08:00) and afternoon (17:00) rush hours, not only the accident counts but also the shares of accidents decrease significantly. Consequently, the accidents are re-distributed throughout the day with a significant increase in share during 19:00.    

\begin{figure*}[ht]
\centering
\includegraphics[width=\textwidth]{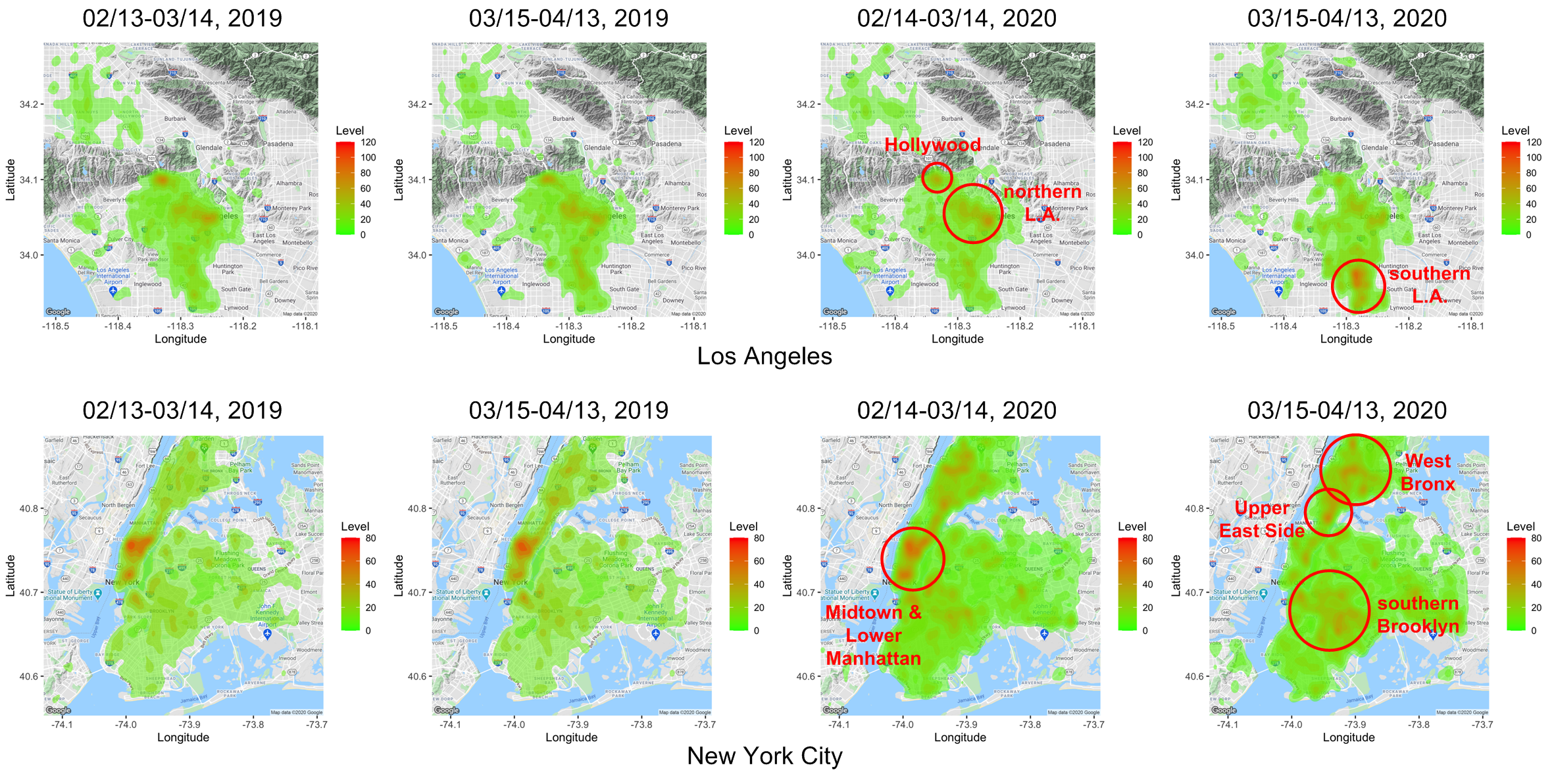}
\vspace*{-.5em}
\caption{\footnotesize{The distributions of traffic accidents in Los Angeles (TOP) and New York City (BOTTOM). Four 30-day analyses are
shown before and after lockdown, in 2019 and 2020, respectively. In Los Angeles (L.A.), the accident hot spots have shifted from the Hollywood area and northern L.A. to southern L.A. In New York City, the accident hot spots have shifted from Midtown and Lower Manhattan to Upper East Side, West Bronx, and southern Brooklyn.}}
\label{fig:temporal_spatial}
\end{figure*}

We have also identified the spatial irregularity of traffic accidents during the pandemic. Fig.~\ref{fig:temporal_spatial} shows the distributions of traffic accidents in Los Angeles and New York City, before and after lockdown. There are two hot spots of traffic accidents in Los Angeles prior to the pandemic with as high as 80 accidents per month: one around the Hollywood area and the other around northern downtown Los Angeles. In contrast, during lockdown, the hot spots have shifted to southern Los Angeles, with the number of traffic accidents increased to more than 110. The distribution shift is also observed in New York City. Prior to the pandemic, the accident hot spots are distributed around Midtown Manhattan and Lower Manhattan, which are shifted to Upper East Side, West Bronx, and southern Brooklyn during the pandemic. 

\vspace*{-1em}
\subsection*{Accident Severity}
Lastly, we have analyzed the changes in traffic accidents of New York City in terms of severity. As severity levels are typically associated with transportation modes, we further divide the accidents into three types: ones without other transportation modes, ones involving pedestrians, and ones involving motorists. While the counts of accidents with and without injuries drop significantly, we find the number of fatal accidents remained the same across all three types under the reduced traffic flows. Another observation is that the share of no-injury accidents increases significantly after lockdown. However, this shift appears to be reversing as we progress longer (e.g., 60 days) into the pandemic. 
% \begin{figure*}
%     \centering
%     \includegraphics[width=.8\textwidth]{./figures/severity.png}
%     \caption{\footnotesize{Change of daily accidents in New York City after lockdown for different severity levels: no injury, injury, and fatality. The accidents are divided into three types: ones without other transportation modes (Left), ones involving pedestrians (Middle), and ones involving motorists (Right).}} 
%     \label{fig:injury}
% \end{figure*} 
\section{Research Description}
\label{sec:research}

% First and foremost, what is the relationship between the road network, community development, demographics, and mobility behaviors? Second, by leveraging the insights from studying the first question, can we better plan, coordinate, and optimize vehicles in different modalities such as human-driven and autonomous to construct safe, efficient, and resilient traffic flows? Lastly, but most importantly, how can we build an intelligent transportation system to promote equity and fairness in our community development?     

Our first research goal is to understand the relationship between the road network, community development, demographics, and mobility behaviors. In our preliminary study, we have analyzed the relationship of reduced traffic flows and traffic accidents. Next step is to bring in road networks and community statistics, and integrate them with our preliminary study. In particular, we believe that the road network is entangled with community development and mobility dynamics, and its features such as network topology and roadway functional classification have intricate relationships with other components in the intelligent transportation system that are not yet fully understood.  So, using the traffic data before and after the pandemic outbreak. We plan to conduct the following research: 1) develop change-point detection algorithm; 2) conduct difference-in-differences analysis; 3) analyze dynamic distribution shifts; and 4) integrate obtained insights with advanced simulation and machine learning techniques to better plan and coordinate CAVs.

% degree, closeness, and betweenness of each state as a node in the traffic flow network. 

% So, in the first thrust, we aim to use both during-pandemic and historical data to study the relationships among the road network infrastructure, various demographic communities, and human mobility activities. We will use the topic of traffic accidents as a starting point, as it is responsible for the highest cost among all transportation-related problems. To be specific, we plan to explore the causality between the road network (e.g., network topology, roadway functional classification) and mobility dynamics (e.g., the change of spatial and temporal traffic accident patterns). We will use frequent pattern analysis to identify critical network and human factors that lead to the distribution shift of traffic accidents associated with different transportation modes. We expect to answer questions such as which demographics experience higher traffic accident rates? When and where do these accidents appear? What are the travel purposes behind these accidents (shopping, work, entertainment, etc)? Understanding these aspects can help design effective policies to fulfill travel demand and reduce traffic accident rates.

\textbf{Change-point detection.} Populace may not react in synchronization with government guidance and policies. So, instead of using the lockdown date from the government, we plan to develop an algorithm to detect change-points in time series data, including traffic data such as  flow and accidents, and pandemic-related data such as confirmed cases. Formally, we consider a non-stationary time series $m=\{m_t\}_{t=1}^T$, which may have abrupt changes at $K$ unknown time steps $1<t_1<t_2<\cdots<t_K<T$. And our goal is to automatically find these unknown time steps via solving the following optimization program:
\begin{equation}\label{eqn:opt}
    \min_{\tau} V(\tau) + \beta K,
\end{equation}
where $\beta$ is the weighting factor; $\tau = \{t_1,t_2,\cdots,t_K\}$ represents the segmentation of the time series. Both $\tau$ and $K$ are unknown and will be identified using our algorithm. $V(\tau)$ is defined as:
\begin{equation}
    V(\tau) = \sum_{k=0}^{K}c(m_{t_k}..m_{t_{k+1}}), 
\end{equation}
where we additionally set $t_0 = 1$ and $t_{K+1} = T$; $c(\cdot)$ is the cost function that measures the similarity of the elements in the time series segment $m_{t_k}..m_{t_{k+1}} = \{m_t\}_{t_k}^{t_{k+1}}$.

% $pen(\tau)$ represents the regularization term, which is used to prevent the time series from being split into too many segments:
% \begin{equation}
%     pen(\tau) = \beta|\tau|,
% \end{equation}
% where $\beta$ is a weighted parameter, and $|\tau|$ is the number of change points. 

\textbf{Difference-in-differences analysis}. In order to statistically test the correlations between road network features, demographics, and traffic behaviors. We need to remove possible seasonal changes in our study to ensure an unbiased analysis. This can be achieved via difference-in-differences (DID) analysis. We will develop our version of DID algorithm to better suit the study of traffic data. As control, pre-pandemic road network, demographic, and traffic data will be used. One way to design the algorithm is to use the following regression model: 
\begin{equation}
    y \sim time*lockdown*x
\end{equation}
where $y$ is the dependent variable; \textit{time} indicates the year; \textit{lockdown} indicates whether a day implements lockdown or not; and $x$ is the independent variable, which could be factors such as age, gender, race, etc. This formula considers not only individual variables but also their two-way and three-way interactions. The interaction coefficients can be used to analyze changes in variables of interest before and after the pandemic outbreak.

\textbf{Dynamic distribution shifts}. To study the potential distribution shifts of various factors, we plan to develop an algorithm based on kernel density estimation. Specifically, we will split the available data regarding the factors of interest in multiple time periods. Again, pre-pandemic data will be used as control. Then, in each time period, we will fit a statistical distribution, e.g.,  the bivariate normal kernel, to pursue the estimation. 

We plan to deliver both qualitative results through visualizing the resulting distributions and quantitative results in the form of statistical comparisons of the distributions. In particular, for the quantitative results, we will conduct a global two-sample test with respect to the integrated squared error (ISE) between the two density functions: 
\begin{equation}\label{eqn:ISE}
ISE = \int (f_1(x) - f_2(x))^2dx,
\end{equation}
where the null hypothesis is $H_0: f_1=f_2$. 
% The test is carried out using the ks package in R.

\textbf{Integration with Simulation and Machine Learning}. After gathering the analysis results from aforementioned studies. We would like to explore the use of connected and autonomous vehicles (CAVs) to improve our transportation systems’ safety, equity, and resilience. Specifically, we plan to simulate city-wide traffic demand, and leverage deep reinforcement learning (DRL) to generate policies for distributing CAVs to meet the traffic demand, while reducing traffic congestion. DRL is a promising technique for large-scale, complex optimization problems, which are challenging for conventional control methods. By unifying human factors such as travel safety and community equity, and system factors such as network resilience as reward signals, we expect DRL to minimize traffic accident rates (an example of safety) and network travel time (an example of resilience), and maximize the number of satisfied trips (an example of equity). We will compare the performance of DRL with traditional vehicle dispatching algorithms under different CAV market penetration rates.

% \subsection*{Work Plan and Milestones}
% \begin{itemize}
%     \item Month 1 (May 15--Jun 15): Download data, pre-processing data, and develop the change-point detection algorithm.
%     \item Month 2 (Jun 16--Jul 15): Develop the DID and distribution shift algorithms and conduct corresponding analyses; Generate quantitative and qualitative results.
%     \item Month 3 (Jul 16--Aug 15): Explore the integration of the acquired insights with advanced simulation and machine learning techniques to plan and coordinate CAVs to improve our transportation system. 
% \end{itemize}

\subsection*{Qualifications}
The PI has extensive experiences on intelligent transportation systems~\cite{Li2017CitySparseITSM,Li2018CityEstIET,Lin2019Compress,Lin2022Attention,Poudel2021Attack,Lin2022GCGRNN,Poudel2022Micro,Shen2022IRL}, traffic simulation~\cite{Wilkie2015Virtual,Li2017CityFlowRecon}, urban mobility~\cite{Lin2021Safety,Wang2021Mobility,Lin2019BikeTRB}, machine learning~\cite{Li2019ADAPS,Shen2021Corruption,Wickman2022SparRL,Villarreal2022AutoJoin}, and multi-agent systems~\cite{Li2015Insects,Li2012Distribution,Li2011Purpose}.  
The PI also co-authored a popular survey paper on traffic simulation~\cite{Chao2020Survey}.

 \setcounter{page}{1}

\bibliography{ref}
\end{document}